\documentstyle[12pt]{article}
 \begin{document}
\def\bq{\begin{equation}}
\def\eq{\end{equation}}
\begin{flushright}
{\sl  December, 1996 }\\
{LPTENS-96/64}\\
{UT-Komaba-96/24}\\
\end{flushright}

\begin{center}
{\large\bf 
FANCY AND FACTS IN THE ($d$ - 2) EXPANSION 
OF NON-LINEAR SIGMA MODELS}
\end{center}

\begin{center}
{\bf E. Br\'ezin$^{a)}$ and S. Hikami$^{b)}$} \end{center}
\vskip 2mm
\begin{center}{$^{a)}$ Laboratoire de Physique Th\'eorique, Ecole Normale
Sup\'erieure}\\ {24 rue Lhomond 75231, Paris Cedex 05, France{\footnote{
Unit\'e propre du Centre National de la Recherche Scientifique, associ\'ee 
\`a l'Ecole Normale Sup\'erieure et \`a l'Universit\'e de Paris-Sud}
} }\\
{$^{b)}$ Department of Pure and Applied Sciences, University of Tokyo}\\
{Meguro-ku, Komaba, Tokyo 153, Japan}\\
\end{center}

\vskip 3mm
\begin{abstract}
 We review the existing results on the scaling dimensions of
operators with more than two derivatives in the non-linear sigma
models. We argue that the speculations on the  relevance of these
operators, and correspondingly on the  breakdown of the $(d-2)$
expansion  for the classical Heisenberg model, or for the
one-parameter scaling theory of localization, are
based on a dubious  mathematical analysis.
\end{abstract}

\vskip 5mm

The low temperature expansion of the $N$-vector model is given by a model of 
interacting Goldstone bosons, known as the non-linear sigma model
\cite{Polyakov,Brezinn}. 
The existence of a zero-temperature fixed point in
dimension two has led to an expansion of the critical properties in 
powers of $(d-2)$ which supplements the familiar expansion in powers
of $(4-d)$ based on the linear Landau-Ginzburg-Wilson model 
\cite{Wilson-Fisher};
linear and non-linear refer here to the representation of the $O(N)$ 
symmetry by the order parameter.

These non-linear sigma models were generalized to situations in which
the order parameter is a mapping of a two-dimensional lattice onto a 
target manifold (a sphere in the case of the $O(N)$ model). These 
generalizations are the basis of the conformal field theory approach to string
theory, and they have been also applied to various problems in condensed matter
physics, in particular to Anderson localization \cite{And}, for which the 
manifold is  the homogeneous space $ O(N,N)/O(N)\times O(N)$ 
\cite{Wegner,BHZ,Hikami}, in
the replica limit in which N goes to zero. This had led to an expansion of the
metal-insulator transition in dimension three in powers of $\epsilon = d-2$,
with $\epsilon=1$ at the end. The numerical success of this expansion 
has been unfortunately in no way comparable to that of the Wilson-Fisher
expansion in powers of $(4-d)$. However for the localization problem, there
is no upper critical dimension from which an expansion similar to the $(4-d)$
 would allow us to ignore the poor numerical accuracy of the $(d-2)$
expansion. There is no other analytic approach to the problem.

In view of these difficulties a growing number of articles,  starting
with the work of Kravtsov, Lerner and Yudson \cite{Kra}, extended later
to the more familiar $O(N)$ model by Wegner \cite{Wegner2}, pointed out to
the possible appearance of new relevant operators, which would of course
drastically change the physics and the results of the model.

In this note we would like to put the calculations which led to the alleged 
breakdown of the $(d-2)$ expansion, in the right perspective. Not that
we have any significant additional calculation to report, but we 
would like the reader to judge the seriousness of the problem by presenting
a summary of the situation as it is known at present.

We shall center our discussion on the $O(N)$ model whose physics is under
much better control, and for which the $4-d$ and $1/N$ expansion, plus
a wealth of numerical experiments leave little doubts on its properties.
Let us start by reminding the reader that the model is defined by a lattice
of unit $N$-vectors with nearest neighbour interactions of the form
${\vec S}(x) \cdot {\vec S}(x + a e_\alpha)$           ,
 in which $e_\alpha$ is a unit vector in the direction $\alpha$. In the long
distance limit we can replace ${\vec S}(x)$ by a continuum field 
and the interaction
by ${\vec S}\cdot \nabla^2 {\vec S}$, with the constraint ${\vec S}^2  =1$;
 (the term involving one
derivative of $\vec S$ cancels with the $x-ae_\alpha$ neighbour). 
The model is then defined by a straight low temperature expansion 
over some ordered state. 
However this expansion is plagued with short-distance divergences,  as in 
any continuum field theory, and these divergences have to be regulated. 
This is slightly tricky in this problem; a straight ultra-violet cut-off 
would not work  since it 
 breaks the $O(N)$ invariance of the model. There are two schemes known 
for regulating these divergences without breaking the rotational invariance,
 the dimensional regularization or a lattice, which means simply 
returning to the low temperature expansion of the original classical
Heisenberg model. Of course these two regularizations lead at the end to
 identical results in the critical long-distance limit \cite{Brezinn}. In 
other words one should be clear at this stage: if one does find a problem 
with this theory it is not simply a problem of some abstract non-linear
field theory, but of the low temperature expansion of the Heisenberg
model itself.

The difficulties which have bothered a number of workers concern the scaling 
behaviour of operators with more derivatives than two. Indeed it is clear 
from the very definition of the model that one could have operators such as 
${\vec S}\cdot ({\nabla^2})^2 {\vec S}$,
 or more complicated ones which would involve derivatives and four
spins for instance, as the ones which would be generated by the interaction of
four spins around a plaquette. In any way we know from Wilson theory \cite
{Wilson-Kogut} that we should allow in the Hamiltonian 
 for any possible $O(N)$ invariant
operator and find the fixed point in a space of an infinite number of 
coupling constants. Note however that we know a priori that there is one
and only one relevant operator (in the absence of a symmetry-breaking operator
such as a magnetic field); indeed it is sufficient, in order
 to reach the critical point
 to vary a single parameter, namely the temperature. If we had more relevant
operators, we would have more "knobs" to adjust before we could reach 
criticality. Numerical experiments, expansions such as the $1/N$, have never 
found anything of the sort: it is sufficient to be at $T_c$ to obtain
an infinite correlation length.

In two dimensions, simple dimensional counting of operators involving $2s$
derivatives of the order parameter ( an even number is needed to make 
a scalar), gives their scaling dimension
\bq\label{1}
        y_s = 2 - 2s                               
\eq
which leads to conclude that operators with more than two derivatives are
 irrelevant. There is no operator for $s=0$ since ${\vec S}^2 =1$; 
therefore we are
 left with 
a single relevant operator, namely ${\vec S}\cdot \nabla^2 {\vec  S}$, 
as expected.

In dimension $d$ two things occur : (i) the canonical dimension is changed 
to $(d-2s)$ (ii) scaling anomalies appear, calculable from  the 
renormalization group theory by expanding in powers of $(d-2)$. 
Wegner's result \cite{Wegner2}, based on a one-loop calculation, gave 
\bq\label{2}
     y_s  = d - 2s + {\epsilon s ( s - 1)\over{N - 2}}  + O(\epsilon^2) 
\eq
 and a recent two-loop calculation by Castilla and Chakravarty \cite{C-C}
gave for large $s$ 
\bq\label{3}
  y_s  = d - 2s + {\epsilon s ( s - 1)\over{N - 2}} 
 + {\epsilon^2} [
{2\over{3}} {s^3\over{(N-2)^2}} + O(s^2)] + O(\epsilon^3)
\eq
 (their calculation is done for finite $s$ as 
well but the result is only 
quoted for large $s$). 
Finally let us quote the last available information which is the $1/N$
expansion of Vasil'ev and Stepanenko \cite{VS} which gives
\bq\label{4}
 y_s  = d - 2s + {4\over{N}} {(d - 1)\Gamma(d - 2)[ s ( s - 1) d ( d - 
3) - 2 s ( d - 2)]\over{ ( 4 - d) \Gamma( 2 - {d\over{2}}) \Gamma
( {d\over{2}}- 1)
\Gamma({d\over{2}} - 2) \Gamma({d\over{2}} + 1)}}
+ O({1\over{N^2}})
\eq
i.e by expanding also in powers of $\epsilon=d-2$
\bq\label{5}
y_s  = d - 2s + {\epsilon s ( s - 1)\over{N}} + {\epsilon^2 s\over{N}} 
+ O( {\epsilon^3\over{N}},{1\over{N^2}})
\eq

We remark here that we could determine 
the two loop result of $y_s$ (\ref{3})  
explicitly by noting that  $y_1 = 1/\nu$, where $\nu$ is a critical exponent for the correlation length, and that $ y_2 = - \omega$, where $\omega$ is an exponent for the 
correction to the scaling. Indeed, (\ref{4}) for $s = 2$ agrees with the 
expression of $1/N$ expansion of the exponent of the correction to the scaling by Ma \cite{Ma}.
From the results of $1/N$ expansion, and from (\ref{3}) the two-loop term may be expressed as 
$({2\over{3}} s^3 + a s^2 + ( N + b ) s )t_c^2$. Since we know that it is 
proportional to $(N - 2)$, for both $s = 1$ and 2, this fixes the two unknown constants $a$ and $b$ to be $a = -2$ and $b = - {2\over{3}}$.

Thus we get
\bq\label{6}
   y_s = d - 2 s + s ( s - 1) t_c + [ {2\over{3}}s ( s - 1) ( s - 2) + 
s ( N - 2) ] t_c^2 + O(t_c^3)
\eq
where $t_c$ is obtained from the zero of the $\beta$ function \cite{Brezinn}.
Putting the expression for $t_c$ in the $\epsilon$ expansion into (\ref{6}), 
we have
\begin{eqnarray}\label{7}
  y_s &=& 2 + \epsilon - 2 s + s ( s - 1) {\epsilon\over{N - 2}}
 \nonumber\\
&+& [ {2\over{3}} s ( s - 1 ) ( s - 2 ) - s ( s - 1 ) + s ( N - 2 )] {\epsilon^2\over{ ( N - 2 )^2}} + O ( \epsilon^3)
\end{eqnarray}

We may recover this result by an alternative method: 
we consider again operators with $2s$ derivatives, but for sigma models on 
the target manifolds $O(N)/O(p)\times O(N - p)$. 
From the isomorphism between $O(5)/O(4)$ 
and $Sp(2)/Sp(1)\times Sp(1)$ \cite{Hik2},
we conclude that 
\bq\label{8}
  y_s = d - 2 s + s ( s - 1 ) t_c + [ {2\over{3}}s ( s - 1 ) ( s - 2 ) + 
( 2 p ( N - p ) - N ) s ] t_c^2 + O(t_c^3)
\eq
using now 
\bq\label{9}
t_c = {\epsilon\over{ N - 2}} - {( 2 p ( N - p) - N) \epsilon^2\over{
( N - 2 )^3}} + O(\epsilon^3)
\eq
we find
\begin{eqnarray}\label{10}
  y_s& =& 2 + \epsilon - 2 s + s ( s - 1 ) {\epsilon\over{N - 2}}
+ [ {2\over{3}} s ( s - 1 ) ( s - 2 )\nonumber\\
&-&{2 p ( N - p ) - N\over{N - 2}}
s ( s - 1 ) + ( 2 p ( N - p) - N) s] {\epsilon^2\over{( N - 2 )^2}} 
+ O(\epsilon^3)
\nonumber\\
\end{eqnarray}
which reduces to (\ref{7}) for $p = 1$.

The results of Wegner and followers led to question the possible relevance of
 the operators with more than two derivatives. Indeed, if in two dimensions
they were irrelevant because $y_s \le -2$ for $s>1$, one sees that the 
anomaly, beeing quadratic in $s$ at first order in $\epsilon$ and cubic in s
at second order, could change this conclusion. Fixing d at physical dimension
 three, and say $N=3$, we see that the $s=2$ operator looks already relevant
 and it is a fortiori true for $s>2$. Wegners's calculation is backed by the
two-loop calculation (3) since the coefficient of $\epsilon^2$ is also 
positive.

This is of course a very strange situation since it leads to the
speculation that there would be an infinite number of relevant
operators, whose coefficients would have to be tuned before one
reaches criticality. On the basis of similar calculations for
generalized sigma  models, Kravtsov et al. \cite{Kra}, were led to
conclude that the one-parameter scaling theory of Anderson et
al. \cite{Abrahams} could be invalid.

For the Heisenberg model there is no experimental or numerical
indication that one could have more than one relevant
operator. Therefore one would conclude that, if these higher
derivatives operators were truly relevant, although there is only one
fixed point at zero temperature in dimension two,  one would have to
find a new fixed point in dimension greater than two  which would
describe the critical properties.  However the $1/N$ expansion, for
instance, does not show any other fixed point, although it
interpolates smoothly between the $4-d$ and $d-2$ expansions. So for
large $N$ at least, this scenario with some new fixed point, would run
against all the existing evidence. 

Therefore let us return to the existing series (\ref{1},\ref{2}); they are
obtained by a standard procedure in which $\epsilon$ is a parameter
which goes to zero first. The previous paradoxical  conclusions are
based however on fixing $\epsilon$ and letting s grow. Clearly it
involves an inversion of limits. To make this point more explicit, let 
us focus on the existing large s information. For large $s$ we can write 
the expansion (\ref{3}) as 
\bq\label{11}
    y_s = d -2s [ 1 - {1\over{2}} x - {1\over{3}} x^2  + O(x^3) ]           
\eq
in which we have defined the parameter
\bq\label{12}
        x = {\epsilon s\over{N-2}}                              
\eq
 Let us consider the function 
\bq\label{13}
        f(x) = 1 - {1\over{2}} x - {1\over{3}} x^2  + O(x^3)            
\eq
For large s, one can speculate  that for
higher orders in $\epsilon$ as well, the result for $y_s$ is of the form
\bq\label{14}
        y_s = 1 -2s f(x)                                
\eq
It is quite possible that for the large s the result has this form and 
we are trying to prove it \cite{BHpre}. 
The sign of $f(x)$, for fixed $\epsilon$, large s, i.e. for large x, is
crucial. Near $x=0$ $f(x)$ is positive, but its asymptotic expansion
at the origin, extrapolated as it is, gives a negative result. However 
we know how dangerous it is to study the sign of a function for large
x based on the knowledge of its asymptotic expansion at small x. Even
if this expansion was convergent, it would presumably have
singularities somewhere, and the small x expansion would give no
information on the sign of the function  beyond the closest
singularity. Furthermore, although there are no rigorous results, it
is more than likely that the expansion in powers of x is only
asymptotic and not convergent for any x. The most naive extrapolation
of $f(x)$ , given the existing data, would be to replace it by a [1,1] 
Pad\'e approximant, i.e. by 
\bq\label{15}
      f(x) = {6-7x\over{6-4x}} + O(x^3)                    
\eq
We certainly do not claim that this is in any way a valid
representation of $f(x)$; however for large x, it is certainly as
trustworthy as the small x-expansion (\ref{13}). Of course the Pad\'e
approximation (\ref{15}) is positive for large x, and the relevance of the
large s-operators becomes even more doubtful.

This discussion does not constitute a proof either in any way. We
wanted simply to stress how unilikely is the picture with an infinite
number of relevant operators, and how weak are the mathematical
assumptions on which these speculations are based.

This does not mean that  the non-linear sigma models are always
perfectly sound. Let us mention some real difficulties. Indeed
consider an $O(N)$- invariant Landau potential $V({\vec \phi}^2)$ which would
develop a minimum away from zero below some temperature, in other
words which leads to a first order transition. In the low temperature
phase of the model, there are are still $(N-1)$ Goldstone bosons,
since the analysis requires simply a broken continuous symmetry, but
not necessarily a second order  transition. In the low temperature
phase of this model, we would write a low temperature expansion, which 
would show a fixed point at non-zero temperature above two
dimensions. Clearly it would be wrong to interpret this fixed point as 
a critical temperature.

The situation may even be a little more subtle. We know of cases in
which the Landau potential gives a mean-field second order
transition, but the fluctuations drive it to first order. This seems
to happen if, for the same $O(N)$-symmetry, the field belongs to the
adjoint representation of the group, rather than to the vector one. It 
means that $\phi$ is an $N\times N$ symmetric, traceless, matrix, which
transforms under the rotation $\omega$ as  $\omega^{T}\phi \omega$. 
 The Landau
potential involves two quartic invariants, namely $({\rm Tr}\phi^2)^2$ and
 ${\rm Tr }\phi^4$. The renormalization group at one-loop, in this space of
two coupling constants, shows at fist order in $4-d$, a runaway
solution to an unstable potential, indicative of a first order
transition. However the analysis of the Goldstone mode of the $O(N)$-
symmetry broken down to $O(N-1)$ depends uniquely upon the Lie algebra
of these two groups, and not on the representation of the group to
which belongs the order parameter. Thus, there again, we would find a
fixed point in the $d-2$ expansion, whereas it is not at all clear
that the model has a second order transition.

Therefore we do not claim that the non-linear sigma models are to be
trusted as other well established field theories such as QED.  Our goal was
simply to spell out that the speculations on their possible breakdown
for the classical  Heisenberg model, or for the one-parameter scaling
theory of localization, are based on an unreasonable mathematical analysis.

We acknowledge the support by CNRS-JSPS cooperative research project, and
by CREST of Japan Science and Technology Corporation.

\end{document}